\documentclass[journal]{IEEEtran}
\usepackage{cite}
\usepackage[tight,footnotesize]{subfigure}
\usepackage[cmex10]{amsmath}
\usepackage{graphicx}
\usepackage{color}

\begin{document}


\title{Multi-rate asynchronous sampling of bandwidth-limited signals}


\author{Alfred Feldster, Yuval P.~Shapira, Moshe Horowitz, Amir Rosenthal, Shlomo Zach, and Lea Singer
\thanks{A.~Feldster, Y.~P.~Shapira, M.~Horowitz and A.~Rosenthal are with the Department
of Electrical Engineering, Technion -- Israel Institute of Technology, Haifa 32000, Israel; e-mail: horowitz@ee.technion.ac.il. S.~Zach and L.~Singer are with the Wales Ltd., 11 Tuval St., Ramat-Gan 52522, Israel.
This paper was also submitted to the IEEE Journal of Lightwave Technology.}}


\maketitle


\begin{abstract}

We demonstrate experimentally an optical system for under-sampling several bandwidth limited signals
with carrier frequencies that are not known apriori that can be located anywhere within
a very broad frequency region
between 0-18 GHz.
The system is based on under-sampling
asynchronously at three different sampling rates. The pulses required for the
under-sampling are generated by a combination of an electrical comb
generator and an electro-absorption modulator. To reduce loss
and improve performance the implementation of the optical system is based on a wavelength
division multiplexing technique. An accurate
reconstruction of both the phase and the amplitude of the signals was obtained when
two chirped signals generated simultaneously were sampled.
\end{abstract}

\section{Introduction}

\IEEEPARstart{I}{n} many applications of radars and communications systems it is
desirable to reconstruct a multi-band sparse signal from its samples.
Such a signal is composed of several narrow-band signals with different carrier frequencies that are
transmitted simultaneously. When the carrier frequencies of the signals
are high compared to the bandwidths of the
signals, it is not cost effective and often it is not feasible
to sample at the assumed Nyquist rate $F_{nyq}$ that is approximately equal to twice the maximum carrier frequency of the signals. It is therefore desirable
to reconstruct the signals by under-sampling; that is to say, from
samples taken at rates lower than the Nyquist rate.

There is a vast literature on theoretical methods to reconstruct signals from under-sampled data
\cite{Kohlenberg,Venkantaramani_Bresler, DO, Moshiko, Feng_Bresler, Lin, Herley}. Most methods are based on a periodic nonuniform sampling (PNS)
scheme. In a PNS scheme $m$ low-rate cosets are chosen out of $L$
cosets of samples obtained from time-uniformly distributed samples
taken at a rate $F$ where $F$ is greater or equal to an assumed  Nyquist
rate $F_{nyq}$ \cite{Venkantaramani_Bresler}. The
sampling rate of each sampling channel is thus $L$ times lower than $F$
and the overall sampling rate is $mF/L$ where $m/L \ll 1$.

In a previous work we have demonstrated theoretically a different
theoretical scheme for reconstructing sparse multi-band signals which we call multi-rate
sampling (MRS)\cite{Asynchronous}.
The scheme entails gathering samples
at $P$ different rates. The number $P$ is small and does not depend on the characteristics of a signal.

The approach is not intended to obtain the total minimum sampling rate.
Rather, it is intended to reconstruct signals accurately with a very
high probability at an overall sampling rate significantly lower
than the assumed Nyquist rate under the constraint of a small number of
channels.

The reconstruction method in Ref.~\cite{Asynchronous} does not rely on synchronization
of different sampling channels.
This significantly reduces hardware requirements.  Moreover, unsynchronized sampling relaxes the stringent
requirement in PNS schemes of a very small timing jitter in the
sampling time of the channels. Simulations indicate that asynchronous multi-rate sampling
reconstruction is robust both to different signal types and to
relatively high noise. A very high reconstruction
probability was obtained ($>98$ \%) amongst real signals comprised of four real signals in which
the bandwidth of each signal was about 100 MHz, and in which the carrier frequencies of the signals
were randomly chosen in the region 0-20 GHz. In each simulation the signal was sampled with three sampling channels that operated at 3.8 GHz, 3.9 GHz, and 4 GHz.

The bandwidth of electronic systems is small compared to the bandwidth
available to analog optical systems. Therefore, with electronic systems, when sampling a
signal whose carrier frequencies can be located in a large
frequency region, the spectrum is divided into small frequency
bands. Each band is down-converted and sampled by a different electronic subsystem.
As a result, the size, the weight and the complexity of electronic
systems limit their applications.

Optical systems are capable
of very high performance under-sampling \cite{Avi}. The carrier
frequency of the input signal can be very high on the order of 20
GHz and the dynamic range of the signals can be as high as 70 dB.
The size, weight, and power consumption of optical systems
make them ideal for under-sampling. The under-sampling operation is based on
shifting the entire spectrum to a low frequency region called the
baseband \cite{Avi},\cite{Joadawlkis}.  The down-converted signal is then sampled using a electronic analog to digital (A/D) converter with a bandwidth that is significantly narrower than the carrier frequency of the signal. A previous work \cite{Avi} has demonstrated the under-sampling and reconstruction of a single narrow-band signal. However, the system required apriori knowledge of the carrier frequency of the signal and this knowledge was used to determine the sampling rate.

In this paper we demonstrate experimentally the sampling and reconstruction of two chirped signals that are generated at the same time. The carrier frequencies of the signals was unknown apriori. The sampling system entails under-sampling the signals asynchronously at three different rates.  The pulses required for under-sampling are generated by a combination of an electrical comb generator and an electro-absorption (EA) modulator. Such an optical pulsed source is simple, small, and consumes low power.
Moreover, the source is robust to changes in environmental conditions.  The combination of the comb generator and the EA modulator that are used instead of two electro-absorption modulators that were used in Ref.~\cite{Avi} results in an increase in the output power of the pulse train by about 10 dB.  The simultaneous sampling at different rates is performed efficiently by using methods based on
a wavelength-division multiplexing (WDM) technique that is used in optical
communication systems. Such a technique avoids loss caused by splitting of the signal between the sampling channels.  This technique also enables use of the same electro-optic modulator for modulating all the sampling channels. Consequently, the frequency dependence of the modulator response curve has the same effect on all sampling channels. We have demonstrated experimentally an accurate reconstruction of two chirped signals, generated simultaneously, each with a width of about 150 MHz. The carrier frequencies of the signals
were not known apriori but were assumed to be within a frequency region of 0-18 GHz. The spurious free dynamic range of the system was 87 dB-Hz$^{2/3}$.

\section{Experimental setup}

\subsection{System description}

\begin{figure}[h]
\begin{center}
\includegraphics[height=7.5cm, angle=270]{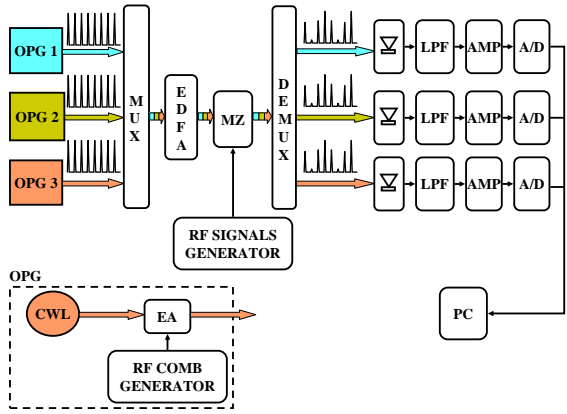}
\end{center}
\caption{ \label{schematics} Schematic description of the system used to under-sample RF signals at
three different sampling rates. Optical pulses are generated using three
optical pulse generator (OPG) units consisting of a CW laser (CWL), a comb
generator and an electro-absorption modulator (EA). By combining three trains of
optical pulses at different optical frequencies using a multiplexer
(MUX) the signal is sampled simultaneously at three different rates. After
modulating the optical pulse trains by the RF signal using a LiNbO$_3$ modulator (MZ), the three
optical modulated pulse trains are separated using an optical demultiplexer (DEMUX), detected
using optical detectors (D), and sampled using a bandwidth limited analog to digital electronic
converters (A/D). AMP and AMP1 are
electrical amplifiers, EDFA is an erbium-doped
optical amplifier, and LPF is an electrical low pass filter.}
\end{figure}

The experimental setup of the system is shown in Fig.~\ref{schematics}. The spectrum of the signals was down-converted to a low frequency region, called baseband, by modulating the amplitude of a train of short optical pulses by the signal \cite{Avi}. Since the reconstruction algorithm requires sampling the signal at three different rates, the down-conversion of the signal was performed separately using three optical pulsed sources that operate at different rates. The wavelengths of the optical pulsed sources were chosen to be different: 1535.04 nm, 1536.61 nm, and 1544.53 nm. Hence, the three optical pulse sources could be added by using an optical multiplexer (MUX) and then be modulated by a single LiNbO$_3$ modulator (MZ). The bandwidth of the modulator used was 40 GHz. The use of a single modulator reduces the sensitivity of the system to the frequency response function of the modulator and simplifies the system design. Since the modulator response
is sensitive to its input polarization we used polarization maintaining (PM) fibers in all of the optical components that were connected before the input end of the modulator. In our experiments we lacked suitable optical multiplexer with PM fibers. Instead, we used two PM couplers to multiplex the three optical signals. This added an additional loss of about 6 dB that can be avoided using a PM multiplexer. The optical losses due to the electro-absorbtion modulator, the electro-optical modulators, the multiplexer and the de-multiplexer, were about 10 dB, 7 dB, 1.5 dB, and 1.5 dB, respectively. To compensate for these losses we used an erbium doped fiber amplifier (EDFA) with a 30 dB gain. After passing through the modulator, the optical wave passed through a demultiplexer that splits the optical signal into three pulse trains of different rates with an amplitude modulated according to the RF signal. Each of the three modulated pulse trains was detected by an optical detector connected to a transimpedance amplifier. The input optical power at the entrance of the detectors was about -4 dBm. The output of the transimpedance amplifiers was then passed through low-pass filters with a cutoff frequency of 2.2 GHz, amplified once again by electrical amplifiers with a gain of 20 dB, and sampled by three electronic analog to digital (A/D) converters. Each of the A/D converters sampled at a rate of 4 Gsamples/sec with 5 effective bits. In each sampling the number of sampled points in each channel was equal to 32,767.

\begin{figure}[htp]
     \centering
     \subfigure[]{
          \label{comb_time}
          \includegraphics[width=7.5cm]{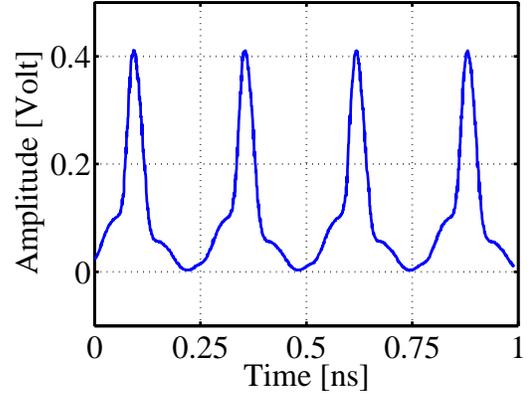}}
     \subfigure[]{
          \label{comb_freq}
          \includegraphics[width=7.5cm]{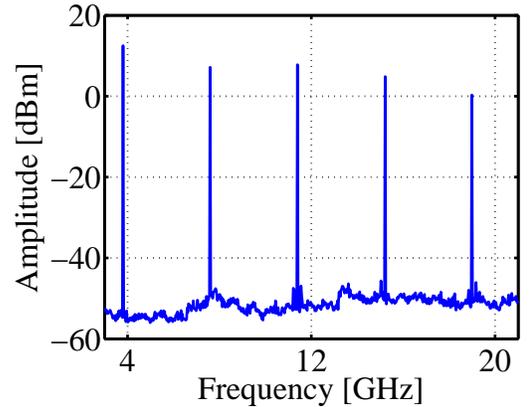}}
    \caption{ \label{comb} Electrical pulses at the output of the comb generator
measured by a sampling oscilloscope and an electrical spectrum
analyzer. The repetition rate of the pulses equals 3.8 GHz.}
\end{figure}

\begin{figure}[htp]
     \centering
     \subfigure[]{
          \label{comb_time}
          \includegraphics[width=7.5cm]{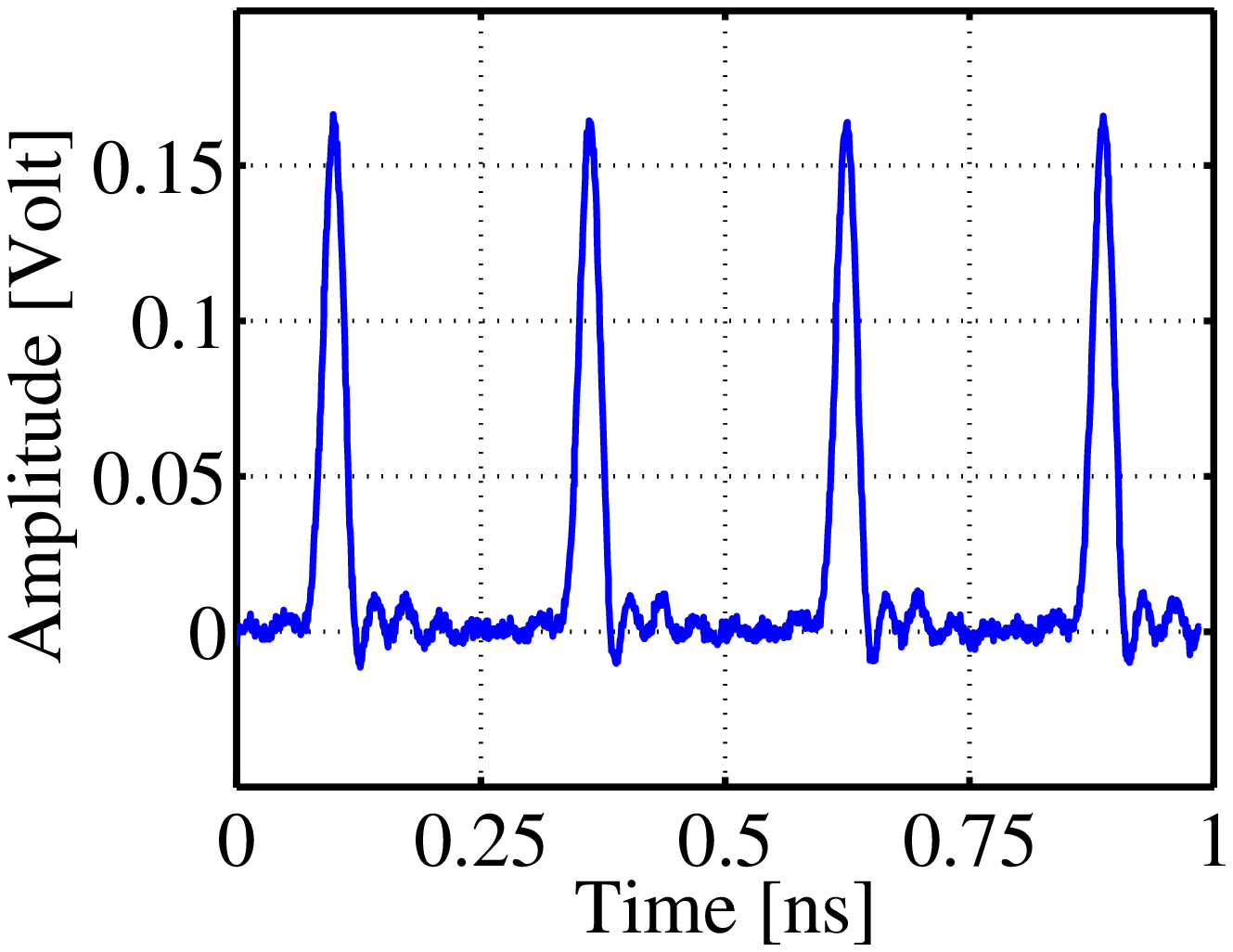}}
     \subfigure[]{
          \label{comb_freq}
          \includegraphics[width=7.5cm]{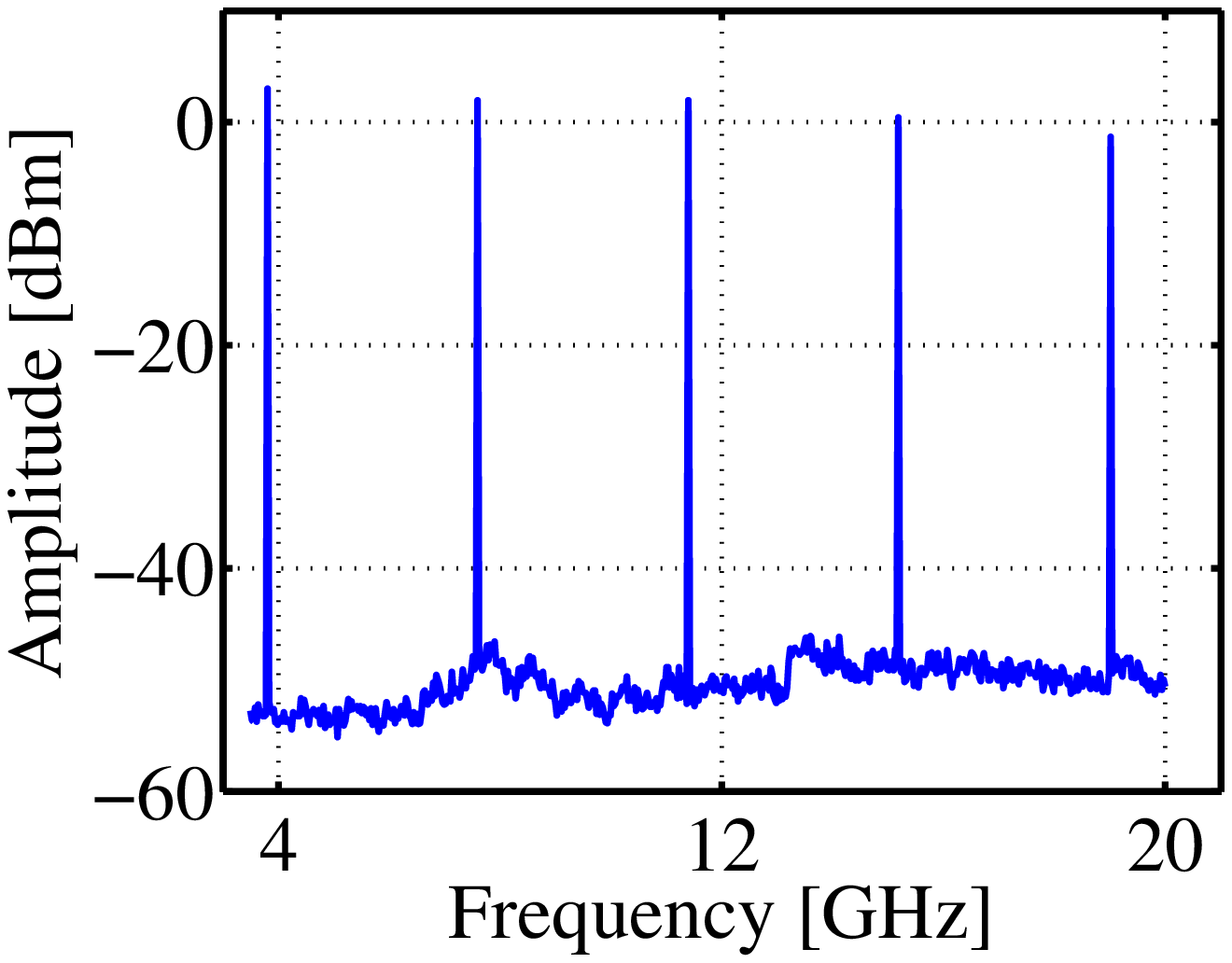}}
    \caption{ \label{opt_comb} Optical pulses at the output of the EA modulator, detected by an optical detector, and measured by a sampling oscilloscope and an electrical spectrum analyzer.
The repetition rate of the pulses equals 3.8 GHz}
\end{figure}

Each pulsed source was implemented by a combination of an EA modulator and a comb generator. The comb generator is based on using a step-recovery diode (SRD) and is used to generate short electrical pulses from a sinusoidal electrical wave. Figure \ref{comb} shows the electrical pulse train at the output of the comb generator.  The train is measured by a sampling oscilloscope and by an RF spectrum analyzer. The measurement shows the existence of strong ripples between the pulses. Moreover, the difference between the harmonics at 3.8 GHz and 19 GHz is 12.3 dB. In Ref.~\cite{Avi} it has been shown that the bandwidth of the system is limited by the bandwidth of the RF spectrum of the optical pulses. Consequently, the pulse train at the output of the comb generator cannot meet the requirement of sampling signals whose carrier frequency may be as high as 18 GHz. To rectify this, we connected the output of the comb generator to the RF input port of an EA modulator with a bandwidth of 40 GHz. The optical input to the modulator was a continuous wave (CW) laser with a power of 11.5 dBm. The nonlinearity of the modulator enabled a shortening of the optical pulse duration as well as a reduction in the intensity of ripples between pulses. Such improvement in the pulses is essential for high performance and high bandwidth sampling. Figure \ref{opt_comb} shows the optical pulses at the output of the EA modulator after conversion to an electric signal by an optical detector with a bandwidth of 40 GHz and measured by an RF spectrum analyzer and a sampling oscilloscope. The pulse duration obtained was about 26 ps. The difference between the RF harmonics at 3.8 GHz and the harmonics at 19 GHz was only 4.3 dB. The average power of each pulse train was about -11 dBm. The repetition rate of the pulses can be controlled easily by changing the frequency of the sinusoidal wave at the input of the comb generator. Using a comb generator and an EA modulator instead of two EA modulators as were used in Ref.~\cite{Avi} enables a reduction in the loss in the pulse source by about 10 dB. We note that the three optical pulse generators in our system were uncorrelated in time.

\subsection{Principle of operation}
The under-sampling is performed in two steps. In the first step the
entire signal spectrum is down-converted to a low frequency region
called baseband by modulating the amplitude of an optical pulse train by the RF signal.
In the second step the down-converted optical signal is
transformed
into an electronic signal that is then sampled by a bandwidth limited electronic A/D
converter. The bandwidth of the electronic A/D converter is significantly narrower than
the maximum carrier frequency of the signals.
The optical under-sampling is performed at three different rates.
The repetition rate of the optical pulse train in each of the
channels is chosen to be different and should be lower than the
sampling rate of the A/D converters to avoid additional aliasing.
Sampling at high rates has a fundamental advantage when applied to
signals contaminated by noise. The spectrum evaluated at a baseband
frequency $f_b$ in a channel that samples at a rate $F$ is the sum of
the spectrum of the original signal at all frequencies $f_b+mF$ that
are located in the overall system operational bandwidth, where $m$ is an integer
\cite{Asynchronous}. Thus, the larger the value of $F$, the fewer
terms contribute to this sum. As a result, sampling at a higher rate
lowers the noise contribution and hence increases the signal to
noise ratio in the baseband. Moreover, our reconstruction
method that is described in Ref.~\cite{Asynchronous} enables a
significant reduction in the required number of sampling channels when the
sampling rate of each channel is increased. The sampling rates that were used in
our experiments were 3.8 GHz, 3.9 GHz, and 4 GHz.


We denote the input wave spectrum at the RF input of the modulator by
$S(\omega)$.
The signal is sampled using an optical pulse train with a temporal pulse shape $p(t)$ and a repetition rate $F$.
The current spectrum $i(f)$ of the optical detector is proportional to\cite{Avi}
\begin{equation} \label{spec} i(f) \propto F
\sum_{n=-\infty}^{\infty} S(f-nF)P(n F),
\end{equation}
where  $P(f)$ is the Fourier transforms of a single optical pulse $p(t)$.
The spectrum at the output of each sampling channel contains replicas of the
original spectrum $S(f)$ shifted by an integer multiple of the repetition rate $F$.
The signal spectrum is therefore down-shifted to a baseband where it can
be sampled using a conventional electronic A/D converter. Because the sampling rate of each
sampling channel is different, the corresponding spectrum in the baseband of each channel is different.

\begin{figure}[h]
\begin{center}
\includegraphics[width=7.5cm]{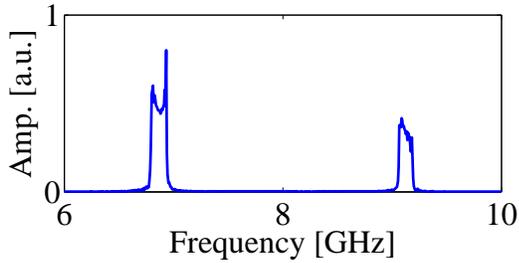}
\end{center}
\caption{ \label{sig_original} Spectrum of the RF signal at the input of the MZ modulator
as measured by an RF spectrum analyzer. The wave contains two chirped signals centered around 6.87 GHz and 9.13 GHz with a bandwidth of 150 MHz and 135 MHz, respectively.}
\end{figure}

\begin{figure}[htp]
     \centering
     \subfigure[]{
          \label{ch3_8}
          \includegraphics[width=7.5cm]{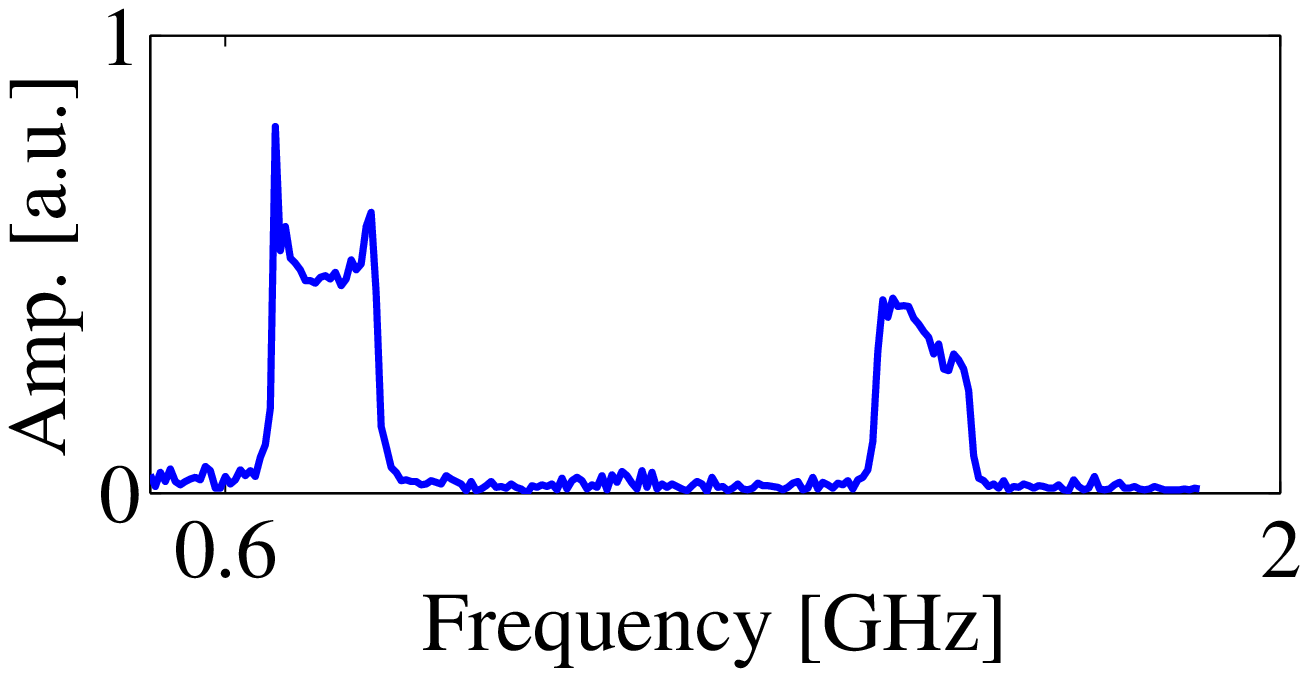}}
     \subfigure[]{
          \label{ch3_9}
          \includegraphics[width=7.5cm]{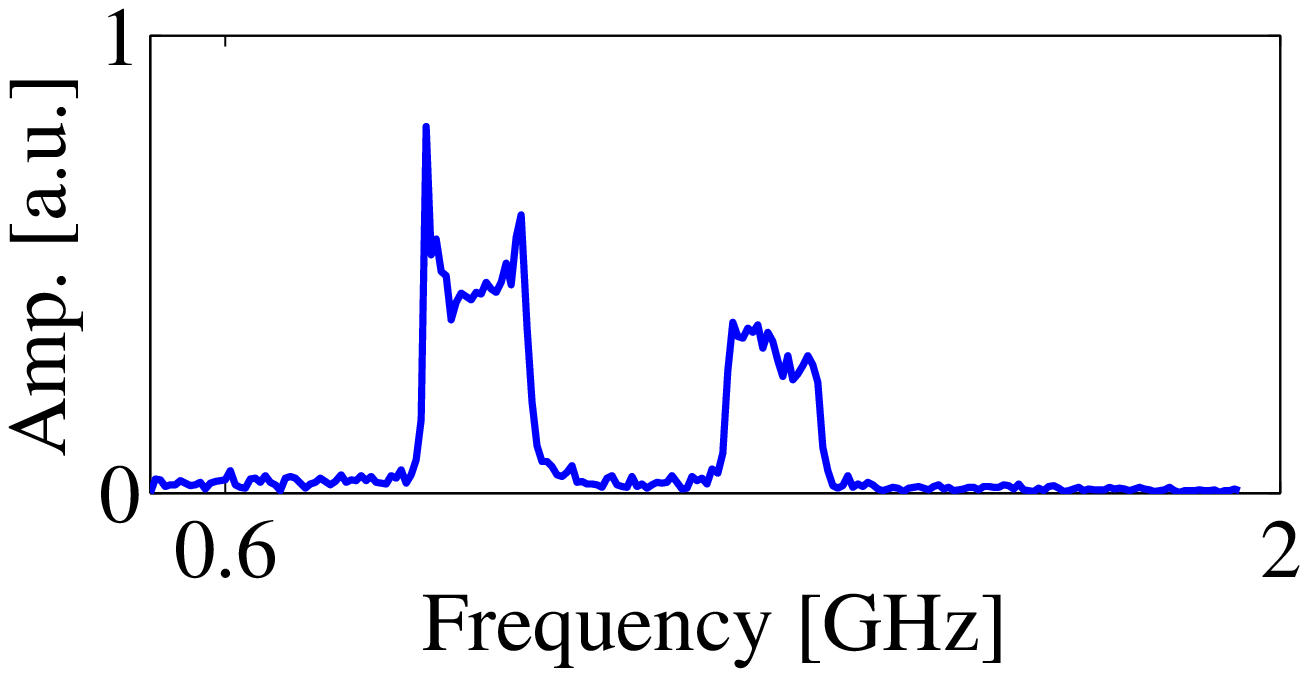}}
     \subfigure[]{
          \label{ch4_0}
          \includegraphics[width=7.5cm]{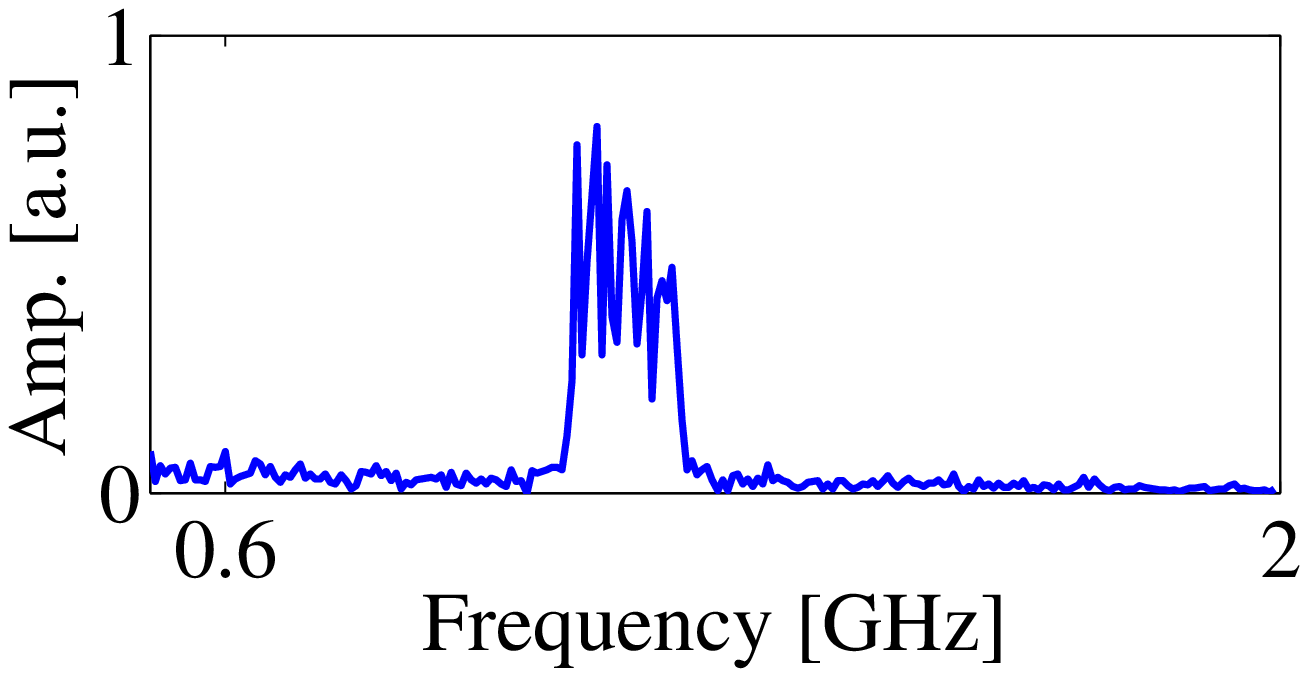}}
    \caption{ \label{samples} Baseband spectrum after sampling using a pulse train with a repetition
    rate of (a) 3.8 GHz, (b) 3.9 GHz, and (c) 4.0
    GHz. The spectrum is the amplitude of the Fourier transform of the sampled data.}
\end{figure}

A signal processing algorithm utilizes the information from the three sampling channels to
reconstruct the signal \cite{Asynchronous}. The algorithm can correctly
reconstruct the signal in almost all cases; even in the cases when aliasing
deteriorates the spectrum in the baseband. The algorithm is based on extracting
a spectrum that minimizes the least square error between the
three down-converted spectra of the reconstructed signal and the spectra measured in the three
sampling channels. The signal can be reconstructed accurately when
each of the frequencies of the signal spectrum is unaliased
at least in one of the sampling channels. Our simulations indicate that
 the multi-rate sampling scheme is robust both to different signal types and to relatively high noise. 

\subsection{Experimental results}

In our experiment we have demonstrated the sampling and the reconstruction of
two chirped signals that are generated simultaneously. Figure \ref{sig_original} shows the spectrum of the input
RF wave. The wave consisted of two chirped pulses with approximate square time profiles. The waves were
generated using a voltage controlled oscillator (VCO). Because the change of
the frequency of the VCO as a function of the input control voltage is not
linear, the chirp obtained did not change linearly in time. The first signal
had a central frequency of 6.87 GHz and a bandwidth of 150 MHz. The second
signal had a carrier frequency of 9.13 GHz and a bandwidth of 133 MHz. The
average RF power of the superposed signals was approximately -14 dBm. The two signals
were generated simultaneously and therefore we have studied the worst case scenario where the two signals
completely overlap on time.
The duration of the combined pulse was 1.35
$\mu$s. The repetition rate of the pulses was 2 kHz. We note that our
simulation results indicate that the sampling system can be used to efficiently sample and reconstruct
more than 4 chirped signals that are transmitted simultaneously \cite{Asynchronous}. However, our source that generated the input signals could
only generate simultaneously two chirped signals.

Mathematically,
since the signals in our experiments are real functions each real signal is composed of two bands.
One band
with a spectrum $S(f)$ is located in the positive frequency region, and
another band with a spectrum $S^*(-f)$ is located in the negative
frequency region \cite{Asynchronous}. Therefore, mathematically, the signal spectrum in our system
is composed of four frequency components. We note that due to sampling each two of the down-converted frequency components can overlap.

Figure \ref{samples} shows the spectra at the output of the three sampling
channels. The spectra were calculated by performing a discrete Fourier transform on the
digital samples from the outputs of the three A/D converters.
We note that after down-converting the signal using a pulse train with a rate of
$F$ the spectrum becomes periodic with a periodicity
of $F$. Moreover, since the signal is real the down-converted spectrum obeys: $i(f)=i^*(-f)$. Therefore,
when sampling at a
rate of $F$, all the information about the down-converted signal is contained in
the frequency region $[0,F/2]$. Figure \ref{ch4_0}
shows that down-converting the signal at a rate of 4 GHz resulted in an
interference in the spectrum at a frequency around 1.1 GHz due to aliasing
effect. The aliasing was between the signal components at $-6.87$ GHz and $9.13$
GHz.

\begin{figure}[h]
\begin{center}
\includegraphics[width=7.5cm]{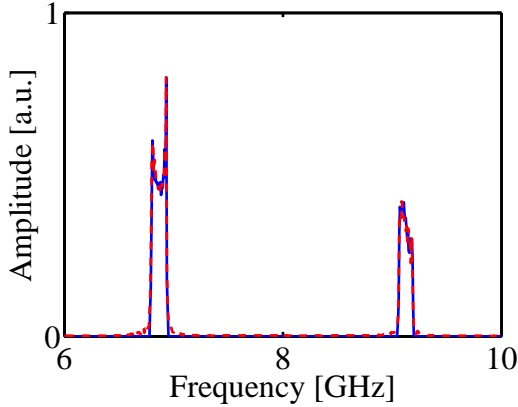}
\end{center}
\caption{ \label{compare} Amplitude of the reconstructed signal spectrum
(blue solid line) compared to the spectrum of the input RF signal that
was measured using a spectrum analyzer (red dotted line).}
\end{figure}

\begin{figure}[h]
\begin{center}
\includegraphics[width=7.5cm]{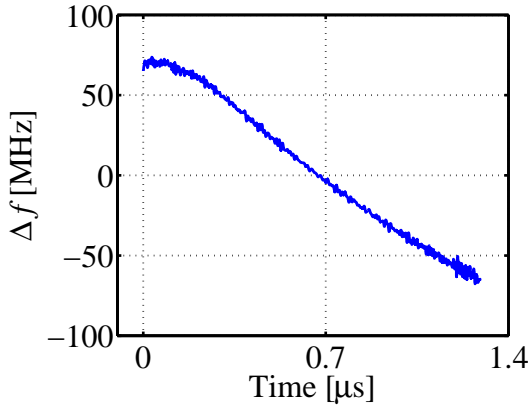}
\end{center}
\caption{ \label{d_phase} Change in the instantaneous frequency $\Delta f = \frac{1}{2\pi}\frac{d\phi}{dt} - f_0$ of the reconstructed signal that is centered around a frequency $f_0 = 6.87$ GHz. $\phi$ is the phase of the reconstructed signal in the time domain.}
\end{figure}

Figure \ref{compare} shows the amplitude of the Fourier transform of the
reconstructed signal. The algorithm for reconstructing the signal from the
three sampled down-converted signals is described in detail in Ref.~\cite{Asynchronous}. For the
comparison, the original spectrum as measured by the RF spectrum analyzer was
added to the figure. The comparison between the original and the reconstructed
signal as shown in Fig.~\ref{compare} shows an excellent quantitative agreement
between the spectrum measured using the analog RF spectrum analyzer and the
reconstructed spectrum. An accurate reconstruction was obtained despite aliasing at
the channel corresponding to a sampling rate of 4 Ghz. Since the
measurement of the RF spectrum analyzer is based on a slow frequency scan,
measuring the whole spectrum requires many RF pulses.
Our system requires only a single RF pulse for measuring the entire spectrum and hence it
can be used in real-time applications. Our system also allows a measurement of the
phase of the input RF signal. This important information can not be obtained
using the RF spectrum analyzer. Figure \ref{d_phase} shows the
instantaneous frequency $i.e.$ the time derivative $d\phi(t)/dt$ of the phase of the
reconstructed signal centered around the frequency $f_0 = 6.87$ GHz. The instantaneous frequency changes in a frequency region of about 142 MHz. This result is in good quantitative agreement with the measured overall signal bandwidth of about 150 MHz. An accurate reconstruction of two signals was obtained in all of our experiments when the carrier frequencies of the two signals were chosen randomly from the frequency region 3-11 GHz. This frequency region was imposed by the bandwidth of our RF sources that were used to generate the signals, and not due to the limitations of the sampling system. We note that in our reconstruction algorithm we assumed that the system bandwidth was 0-20 GHz.
\begin{figure}[h]
\begin{center}
\includegraphics[width=7.5cm]{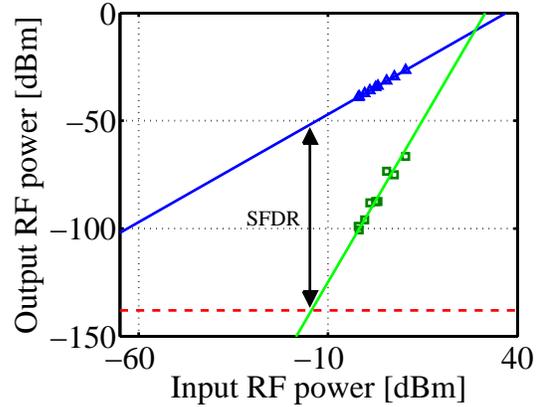}
\end{center}
\caption{ \label{SFDR} Measurement of the spurious free dynamic range (SFDR) of
the sampling channel obtained using a pulse train with a repetition rate of 3.9
GHz. The triangles show the measured input power of each of the two CW RF
signals at frequencies 6.197 GHz and 6.198 GHz. The squares show the
corresponding measured power of the third order products at the output of the
optical detector (see Fig.\ref{schematics}). The SFDR as shown in the figure
was equal to 87 dB-Hz$^{2/3}$.}
\end{figure}

Figure \ref{SFDR} summarizes the measurement of the the spurious free dynamic
range of the system. In this measurement the electro--optical modulator was
driven by two continuous wave (CW) RF signals of the same power at adjacent
frequencies of 6.197 GHz and 6.198 GHz. The power of the two CW
electrical signals was gradually changed between -2 dBm and
10.5 dBm. For every value of the input power, the power of the third order
products at frequencies 6.196 and 6.199 GHz was measured at the output of each
of the three photo-detectors. The obtained spurious free dynamic range of each
of the channels was about 87 dB-Hz$^{2/3}$. We believe that by using narrow
optical filters as well as optical multiplexer instead of couplers we will be
able to reduce significantly the noise in the system.

The bandwidth of the system was measured by scanning the frequency of a CW RF
source that was connected to the system input. The power of the down-converted
signal was measured as a function of input signal
frequency. The 3-dB bandwidth was about 17.8 GHz. This result is in
agreement with the bandwidth of RF spectrum of the optical pulses shown in
Fig.~\ref{opt_comb} as expected by theory \cite{Avi}.
Therefore, shortening the optical pulses will increase the system bandwidth.


\section{Conclusion}

We have demonstrated a new optical system for under-sampling sparse multi-band
signals. Such signals are composed of several bandwidth limited signals. The
carrier frequencies of the signals are not known apriori and can be located
anywhere within a very broad frequency region (0-18 GHz). The system is based on under-sampling
asynchronously the signals at three different rates. The under-sampling is
performed by down-converting all of the signal spectra to a low frequency region
called the baseband. The baseband is then sampled using electronic
analog-to-digital converters with a bandwidth that is significantly narrower than
the maximum carrier frequency of the input signals. By separating the frequency down-conversion and the
analog to digital conversion operations it becomes possible to use a
single frequency resolution that is common to all of the sampling channels.
This facilitates a
robust reconstruction algorithm that is used to detect and reconstruct the phase and
the amplitude of the signals. The reconstruction method does not rely on
synchronization between different sampling channels. This significantly reduces
hardware requirements. Because the entire frequency region of the signal is
down-converted to a baseband, aliasing may cause a deterioration in the
down-converted spectrum. However, when the sum of the signals forms a sparse
wave, a very high theoretical successful reconstruction percentage ( $> 98 \%$) can be obtained using our
reconstruction algorithm. Our system uses standard optical components that are used in optical communication systems. The optical system is
robust against changes in environmental conditions. It is a turn-key system
that does not require any adjustments prior to the start of its operation. The
weight, power consumption, dynamic range, and bandwidth of the optical
system are significantly superior to those of electronic systems capable of
meeting similar requirements. The performance of the reconstruction algorithm
can be further enhanced by synchronizing the three sampling channels.
With such a synchronization aliasing can often be resolved by solving a set of linear
equations.


\begin{IEEEbiography}[]{Yuval P.~Shapira}
was born in Russia, in 1980. He has received his B.Sc. degree (2006) and his M.Sc. degree (2008), summa-cum-laude, from the Technion - Israel Institute of Technology. He is currently working towards his Ph.D. degree in electrical engineering, at the Technion.
His research interests include fiber optics and non-linear effects in fiber Bragg gratings.
\end{IEEEbiography}

\end{document}